\documentclass[prc,aps,nofootinbib,showkeys,showpacs,twocolumn,floatfix]{revtex4} 
\usepackage{epsfig}
\usepackage{graphicx}
\usepackage{ulem}

\begin{document}
\title{On the description of two-particle transfer in superfluid systems}

\author{ Danilo Gambacurta and Denis Lacroix} 
\affiliation{GANIL, CEA and CNRS/IN2P3, Bo\^ite Postale 55027, 14076 Caen Cedex, France}

\begin{abstract}
Exact results of pair transfer probabilities for the Richardson model with
equidistant or random level spacing are presented. The results are then compared
either to  particle-particle random phase approximation (ppRPA) in the normal phase or quasi-particle
random phase approximation (QRPA) in the superfluid phase. We show that both 
ppRPA and QRPA are globally well reproducing the exact case although some differences are seen in the 
superfluid case. In particular the QRPA overestimates the pair transfer probabilities
to excited states in the vicinity of the normal-superfluid phase
transition, which might explain the difficult in detecting collective
pairing phenomena as for example the Giant Pairing Vibration.
The shortcoming of QRPA can be traced back to the breaking of particle number that
is used to incorporate pairing. A method, based on direct diagonalization of the
Hamiltonian in the space of two quasi-particle projected onto good particle
number is shown to improve the description of pair transfer probabilities in superfluid
systems.      
\end{abstract}

\maketitle

\section{Introduction}
The importance of pairing correlations in nuclear systems has been established 
in several aspects, binding energies of nuclei, odd-even  effects, superfluid phenomena
and pair transfer mechanisms, to mention just a few. 
However, despite the fact
that pairing is anticipated to play a significant role in the pair transfer
process, the existence of collective effects, like Giant Pair Vibration (GPV)
\cite{Bro73, Oer01,Pot09,Kha04,Kha09,Ave08} leading to an increase in the pair
transfer from a superfluid nuclei still challenges the experimental nuclear
physics \cite{Mou11}. On the theoretical side,
mean-field methods based on Hartree-Fock-Bogolyubov (HFB) sometimes augmented by
Quasi-particle Random Phase Approximation (QRPA) have been used to predict pair
transfer probabilities either from ground state to ground state or from ground
state  to excited states  \cite{Dob96,Shi11,Pot11,Pll11,Gra12}. A common
conclusion of most of these studies is the sensitivity of one- or two-nucleon
transfer process to the internal topology of pairing in nuclei. In the present
work, exact results of pair transfer probability are obtained for the Richardson
model \cite{Ric64} consisting in set of single-particle levels interacting through a pure
pairing interaction. This model can be seen as the valence space of the last
occupied level in nuclei where nucleons can be either added (pick-up reactions)
or removed (stripping reactions). The possibility to perform exact calculations
open new perspectives to understand the pair transfer process and provides a
benchmark for approximate treatments. In the following, we first discuss the
pair transfer mechanism from a general point of view and estimate pair transfer
probabilities in the pairing model.  

We are interested here in a process where two particles 
are either added or removed in a system that is initially in its ground state formed of N particles. 
In the following, we denote by $| \nu, A \rangle $
respectively the eigenstates of the systems with $A$ particles associated 
to the set of energies $E^A_\nu$ and, by convention, $\nu = 0$ is taken for ground state.
During its evolution, the system wave-function can be decomposed 
as  \cite{Rip69}
\begin{eqnarray}
 | \Psi (t) \rangle & = &e^{-itE^N_0/\hbar}  \left\{ \sum_\nu c^N_\nu e^{-it(E^N_\nu - E^N_0)/\hbar}| \nu, N \rangle \right.  \nonumber \\
&&\phantom{e^{-itE^N_0}  }+ \sum_\nu c^{N-2}_\nu e^{-it(E^{N-2}_\nu - E^N_0)/\hbar}| \nu,N -2 \rangle  \nonumber \\
&&\phantom{e^{-itE^N_0}  }+ \left. \sum_\nu c^{N+2}_\nu e^{-it(E^{N+2}_\nu - E^N_0)/\hbar}| \nu, N +2 \rangle \right\} \nonumber
\end{eqnarray}
where the first line describes the possibility that the system remains in its
ground state or in one of its excited states without changing its particle number.
Second (resp. third) line contains the information on the removal and/or
addition process.  The explicit form of the coefficients $c^A_\nu$ depends on
the physical process under interest, like the stripping or pick-up reactions in
nuclear physics.
On the theoretical side, information of these processes can be obtained by studying the small amplitude 
response of the system 
to an external field $\hat T$ that changes the particle number by two units.  Then, information on the transfer reduces to the knowledge of the response 
function, given by:
\begin{eqnarray}
S(E)&=&  \sum_\nu  | \langle N+2, \nu | \hat T | N,0\rangle |^2 \delta \left(E -\Delta E^{N+2}_\nu \right) \nonumber \\
&+& \sum_\nu  | \langle N-2, \nu | \hat T | N,0\rangle |^2 \delta \left(E -\Delta E^{N-2}_\nu \right)  \label{eq:strength} \\
 & \equiv & S^{\rm Add}(E) + S^{\rm Rem}(E) \nonumber 
\end{eqnarray}
where $\Delta E^{N\pm2}_\nu = E^{N\pm2}_\nu - E^{N}_0$. In the following, $S^{\rm Add}(E)$ (resp. $S^{\rm Rem}(E)$)
will be referred to addition (removal) strength function.  
A common choice of $\hat T$ \cite{Ave08} to excite pairing modes is
\begin{eqnarray}
\hat T & = & \sum_{i} (T_{i \bar i} a^\dagger_i a^\dagger_{\bar i} + T^*_{i \bar i }    a_{\bar i} a_i )
\label{eq:Top}
\end{eqnarray}
where $a^\dagger_ia^\dagger_{\bar i}$ corresponds to creation operators of a pair of time-reversed 
single-particle states. In the present article, we are interested in a physical process where two-particles are either added or removed.
In that case, it is more suited to consider directly the non-Hermitian addition (resp. removal) transition operator, denoted by $\hat T^{\rm Add}$ (resp. $\hat T^{\rm Rem}$) 
defined through
\begin{eqnarray}
\hat T^{\rm Add} & = & \sum_{i} T_{i \bar i} a^\dagger_i a^\dagger_{\bar i}, ~~~\hat T^{\rm Rem} =  \sum_{i} T^*_{i \bar i }    a_{\bar i} a_i 
\label{eq:Topadd}
\end{eqnarray}
respectively associated to $S^{\rm Add}(E)$ and $S^{\rm Rem}(E)$.
From the expression of the strength,  we see that the understanding 
of the two-particle transfer passes through a good knowledge of the spectroscopy of initial and final 
states as well as of the capacity to provide the addition or removal probabilities 
defined here as:
\begin{eqnarray}
P^{\rm Add}_{\nu} & = & | \langle N+2, \nu | \hat T | N,0\rangle |^2  ,\label{eq:add} \\
P^{\rm Rem}_{ \nu} & = & | \langle N-2, \nu | \hat T | N,0\rangle |^2 .  \label{eq:rem}
\end{eqnarray}
If the many-body problem can be solved exactly, such quantities as well as the exact 
eigenvalues of the Hamiltonian can be used to have a precise estimate of the strength function
(\ref{eq:strength}). In most realistic situations, such treatment is impossible and educated guess should be 
used. In the present work, we are interested in an initial system where pairing correlation plays a role.  
This case has been first considered in refs. \cite{Bes66,Bro68,Bro73} leading to
the concept of pairing vibration, i.e. a coherent excitation of pairs of
particles that is expected to show up in the enhancement of pair transfer
probabilities. The standard way to incorporate pairing correlation is to use the
BCS or HFB approach as a starting point
\cite{Rin80,Bla86}. This technique is indeed standardly used in nuclear physics
either to estimate the transfer from ground state to ground state
\cite{Mat10,Shi11,Gra12} or from ground state to excited states. 
In the latter case, the response is obtained using QRPA \cite{Kha04,Kha09,Pll11} or its time-dependent version \cite{Ave08}. 
For a comprehensive introduction, please refer to \cite{Bri05}. 

With the increase of computational powers, it is possible nowadays to study
exactly pair transfer in schematic model that approaches realistic situations
and to quantify the predictive power of mean-field based approaches. In the
present work, we first present exact results of pair transfer probabilities for
the Richardson model with equidistant or random level spacing. The exact results are then used 
to benchmark standard approaches, namely ppRPA and QRPA.    

\section{Exact description in the Richardson model}

A system of $\Omega$
doubly-degenerated single-particle levels interacting via a pairing force with parameter $G$ is considered.  
The  Hamiltonian is given by \cite{Ric64}
\begin{equation}
H=\sum_{i=1}^{\Omega}\epsilon_{i} \hat N_{i}-G\sum_{i,j=1}^{\Omega}
\hat P_{i}^{\dagger} \hat P_{j}~,
\label{H}
\end{equation}
where the particle-number operator $\hat N_{i}$ and pair creation/annihilation operators 
$\hat P_{i}^{\dagger}$, $\hat P_{i}$ are given by
\begin{equation}
\hat N_{i}=a_{i}^{\dagger}a_{i}+a_{\bar i}^{\dagger}a_{\bar i}~, 
\hspace{5mm} 
\hat P_{i}^{\dagger}=a_{i}^{\dagger}a_{\bar i}^{\dagger}~,~~~~
\hat P_{i}=(\hat P_{i}^{\dagger})^{\dagger}~.
\label{N&P}
\end{equation}
\begin{figure}[htbp] 
\includegraphics[width=8cm.]{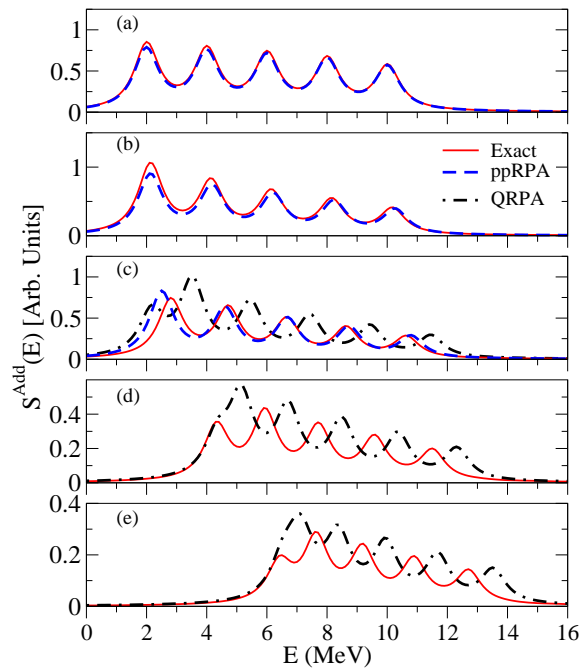} 
\caption{ (color online) Addition strength function as a function of the excitation energy $E$ for different pairing strength G obtained using exact eigenvalues and eigenstates in Eq. (\ref{eq:strength}) (red solid line) for a system of $N=8$ particles to $N=10$ in the case of $\Omega=10$ equally spaced levels. 
From top to bottom: (a) $G/\Delta \varepsilon= 0.1$ , (b) $0.3$, (c) $0.5$, (d) $0.7$ and (e) $0.9$ are shown.  The results of the ppRPA (blue dashed line) and QRPA (black dot-dashed line)
are also presented. Note that in panel (c), the result of ppRPA obtained at the collapse point $G=0.48 \Delta \varepsilon$  is also reported. The QRPA are shown only above the BCS threshold $G=0.33 \Delta \varepsilon$. }
\label{fig1:pair} 
\end{figure} 

\begin{figure}[htbp] 

\includegraphics[width=8cm.]{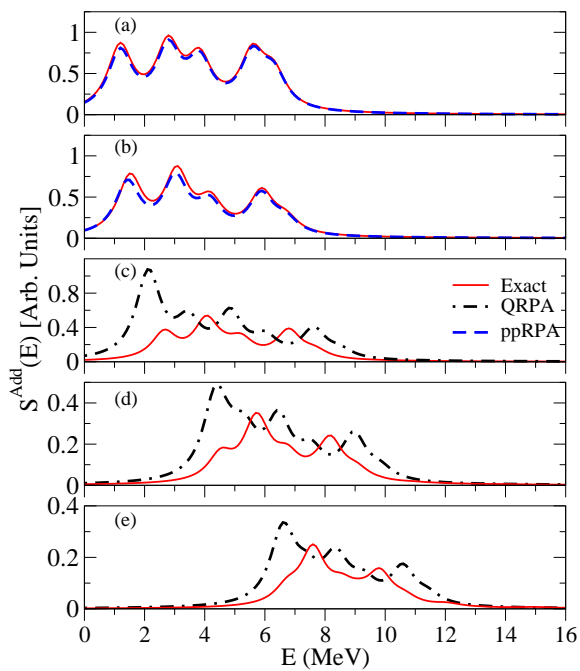} 
\caption{ (color online) Same as figure \ref{fig1:pair} obtained for randomly spaced single-particle energies. 
See the text for more details. The QRPA are shown only above the BCS threshold $G=0.44$ MeV. } 
\label{fig2:pair} 
\end{figure}

For not too large model space $\Omega$, exact solutions can be obtained using standard 
diagonalization techniques in subspace of given seniority \cite{Zel03}.
As a test case here, a system of $N=8$ particles with $\Omega =10$ doubly
degenerated levels is considered here. 
Assuming a simple transition operator (\ref{eq:Top}) with $T_{i \bar i}=1$ for all pairs, illustrations of addition strength function obtained for the Richardson model 
are shown in Fig. $\ref{fig1:pair}$, in full (red) line, for equidistant level spacing, i.e. 
$\epsilon_{i} = i \Delta \varepsilon$ ($i=1$, $\Omega$),
for different values of the pairing interaction and $\Delta \varepsilon= 1$ MeV. 
In the following the excitation energies will be calculated with respect
to the ground state of the system with $N=10$ particles consistently in all the theories.
The presented result is exact in the sense that the exact eigenvalues and 
eigenstates of the system with $N=8$ and $N=10$ have been used to compute Eq. (\ref{eq:strength}). Note that in the present work, we are
mainly  interested in the transition from ground state to excited states of the $N+2$ nucleus and the contribution of the ground state to ground state 
transfer has been omitted in the figure. Moreover, in order to make simpler the comparison between
different results, we have folded the discrete spectra with a Lorentzian function with a width of 1 MeV.
For completeness, we also show in Fig.  \ref{fig2:pair} similar results obtained with randomly spaced system
whose single-particle energy are $0.551$, $2.176$, $4.033$, $5.142$, $6.444$, $7.029$, $7.827$, $8.343$, $9.226$ and $9.571$ in MeV units. 
This situation could be considered closer to realistic cases.

In the exact calculations,  we observe a shift of the excitation spectrum
toward higher energies  as the pairing strength $G$  increases 
while the pair transfer probabilities of the excited states get smaller. On the contrary, as it will be 
shown below, the pair transfer probability from ground state to ground state increases as $G$ increases. 

\section{ppRPA vs QRPA approaches to pair transfer}

As already mentioned, in most cases, exact evaluation of the pair transfer probabilities cannot be 
performed and approximations for the many-body states are necessary.  The most common strategy 
used in nuclear physics is to first apply the HFB or BCS theory  and minimize 
the energy in the Hilbert space of quasi-particle vacuum imposing a mean particle number equal to $N$. 
This leads to an approximation for $| 0, N \rangle$. Using standard notations, the quasi-particle vacuum
is given by
\begin{eqnarray}
| 0, N \rangle   \simeq | 0, {\rm QP} \rangle = \prod_i \left[ \alpha_i^\dagger \alpha_{\bar i}^\dagger\right]  | - \rangle
\end{eqnarray}
where $| - \rangle$ is the quasi-particle vacuum and $\alpha_i$ is the quasi-particle annihilation operator defined through the simple Bogolyubov transformation:
\begin{eqnarray}
\alpha_i & = & U_i a_i - V_i a^\dagger_{\bar i} \\
\alpha_{\bar i} & = & U_i a_{\bar i}  + V_i a^\dagger_{i} .
\end{eqnarray}
In the HFB approach, the above transformation automatically implies that the single-particle basis 
identifies with the canonical basis.

\begin{figure}[htbp] 
\includegraphics[width=8cm.]{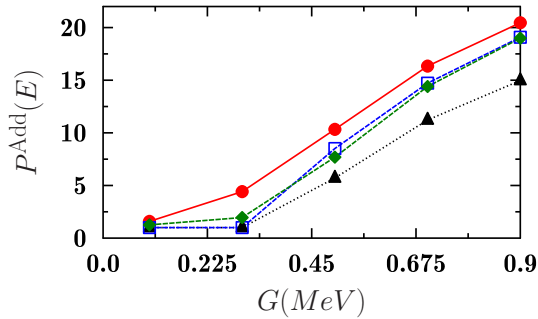} 
\caption{ (color online) Illustration of the application of mean-field theory to provide estimate of ground state to ground state 
addition pair probability. The exact result (red circles), BCS (black triangles) and BCS projected on good particle number (blue open squares)
are shown here for the equidistant single-particle level case. The result of the P-QTDA$^{\rm (GS)}$ is also presented with green filled diamond.} 
\label{fig2bis:pair} 
\end{figure}

In the present model, the Hamiltonian (\ref{H}) is already written in the canonical basis for the HFB theory. 
This theory is particularly suitable to provide estimates for ground state (GS) to GS transfer probabilities \cite{Shi11,Gra12}. Since  we want to compare with exact results, at the BCS/HFB and QRPA level 
the contribution to the particle-hole channel of the pairing interaction is taken into 
account.
The result of the HFB theory for this probability is shown in figure \ref{fig2bis:pair} and compared to the exact solution. Note that
below the pairing threshold (denoted by
$G_{\rm cr}$  and equal to 0.33 MeV and 0.44 MeV for the  $8$ particles system in the equally spaced and random
spaced case, respectively), 
the HFB reduces to HF and probability is $1$. 
As illustrated in this figure, while the HF theory (not shown) would have failed to reproduce the exact probabilities, the HFB framework 
gives estimations that are already rather close to  the exact ones in the superfluid regime. 

Due to the absence of residual coupling between quasi-particle excitations, it is known that HFB alone cannot properly 
describe excited states.  
Then, linear response theory including possible particle-particle (pp),
hole-hole (hh), or particle-hole (ph) excitations is applied to describe
excited states, and then transfer probabilities, within the QRPA approach. In QRPA, the excited
states, denoted by $| \nu \rangle$, are obtained by considering coherent
superposition of 2 quasi-particle (2QP) excitations. This leads to
\begin{eqnarray}
| \nu \rangle &=&  Q_\nu^{\dagger} | 0 \rangle,  \label{eq:nuqrpa}
\end{eqnarray}  
where  $Q_\nu^{\dagger}$ are QRPA phonons written as  
\begin{equation}
Q_\nu^{\dagger}=\sum_i(X_j^\nu\alpha^\dagger_i \alpha^\dagger_{\bar i}-Y_j^\nu  \alpha_{\bar i}\alpha_i),
\hspace{5mm} Q_\nu = (Q_\nu^{\dagger})^{\dagger}, \label{eq:qrpaphonon}
\end{equation}
while $| 0 \rangle$ is the phonon vacuum, defined through the conditions $Q_\nu|0\rangle=0$. In practice, 
the components $X^\nu$ and $Y^\nu$ as well as the energies of the excited states $\omega_\nu$ 
are deduced by solving the QRPA eigenvalue problem \cite{Row10}. Since these techniques are 
rather standard \cite{Rin80}, we only recall here the expressions of the pair transfer probability. 

In QRPA, the addition transition probability is given by \begin{eqnarray}
 P^{\rm Add}_\nu= |\langle 0|\hat T^{\rm Add}| \nu\rangle|^2  =|\sum_{i}(V^2_iX_{i}^{(\nu)} -U^2_iY_{i}^{(\nu)})|^2.
\end{eqnarray}

It is well known that,
the first solution of the QRPA equations corresponds to the spurious
mode, which is then not considered in the evaluation of the strength function.

In the weak coupling limit, below a certain threshold value of $G$ denoted by
$G_{\rm cr}$, the minimization of
the energy in HFB identifies to the Hartree-Fock approach with no pairing. Then
the mean-field vacuum is a pure Slater determinant where the lowest hole states
are occupied. In this case, labeling by $h$ the hole state ($V_h$ = 1, $U_h=0$)
and $p$ ($V_p = 1$, $U_p=1$) the particle states associated to this vacuum, the
excited states are described by using the particle-particle RPA (ppRPA) where the phonon creation operators
(\ref{eq:qrpaphonon}) can be written as 
\begin{eqnarray}
Q_\nu^{\dagger}&=&\sum_{p}
X_{p}^{\nu}a^\dagger_p a^\dagger_{\bar p} +\sum_{h}Y_{h}^{\nu}
a^\dagger_h a^\dagger_{\bar h},
\end{eqnarray}
while the addition probability simply writes:
\begin{eqnarray}
P^{\rm Add}_\nu &=&
 |\sum_{p}X_{p}^{\nu}-\sum_{h}Y_{h}^{\nu}|^2.
 \end{eqnarray}
Note that in the ppRPA case, contrary to the QRPA, the $U(1)$ symmetry associated to particle number conservation 
is not broken. Explicit forms of the ppRPA and QRPA equations for the model considered here can be found in Refs. \cite{Gam06, Dan06,Dan07} 

In Figs \ref{fig1:pair} and \ref{fig2:pair}, the ppRPA (dashed line) and QRPA (dot-dashed line) 
are compared to the exact results.
  Above a given threshold $G_{cr}^{RPA}$, ppRPA collapses and leads to imaginary
energies making not possible a direct comparison with the exact and QRPA results.
However, in Fig. \ref{fig1:pair} we show in the panel (c) corresponding to 
a pairing strength $G=0.5 MeV$ the ppRPA results obtained at the collapse point,
i.e.  $G_{cr}^{RPA}=0.48$, in order to show that its description is still
reasonable even in the superfluid phase.
 
 From these comparisons, the following conclusions can be drawn: (i) The ppRPA
does reproduce perfectly the exact results (energies and probabilities) in the
normal phase. (ii) The QRPA provides a global reproduction of the pair transfer probabilities 
in the superfluid phase. In particular,  the threshold in energy that is directly related to pairing 
correlation is properly accounted for. It is worth to mention, that such a threshold can only be described 
in a mean-field theory by breaking particle number symmetry.  Finally, it is also clearly seen 
that some differences exist between the exact and QRPA. in general QRPA leads to peaks in the strength that are at slightly higher energies
compared to the exact solution while probabilities are slightly overestimated. 
These differences are even stronger in the random space case and can stem from different origins. 
First, while part of the four quasi-particles (4QP) are accounted for 
in the QRPA ground state correlations, the complete inclusion of 4QP excitations is known to modify excited state energy spectrum \cite{Rin80,Lac12}.
Second, a systematic error exists due to the breaking of particle number. Panel (c) of Fig. \ref{fig1:pair} illustrated that when 
ppRPA is applicable in the superfluid phase $G\leq0.48$ MeV, it gives a better
agreement with the exact result compared to QRPA. Since ppRPA is a particle number conserving theory, this is a first indication that
the breaking of $U(1)$ symmetry might pollute the QRPA predictions.    

\subsection{Role of particle number in the estimation of pair transfer probabilities}

\begin{figure}[htbp] 
\includegraphics[width=8cm.]{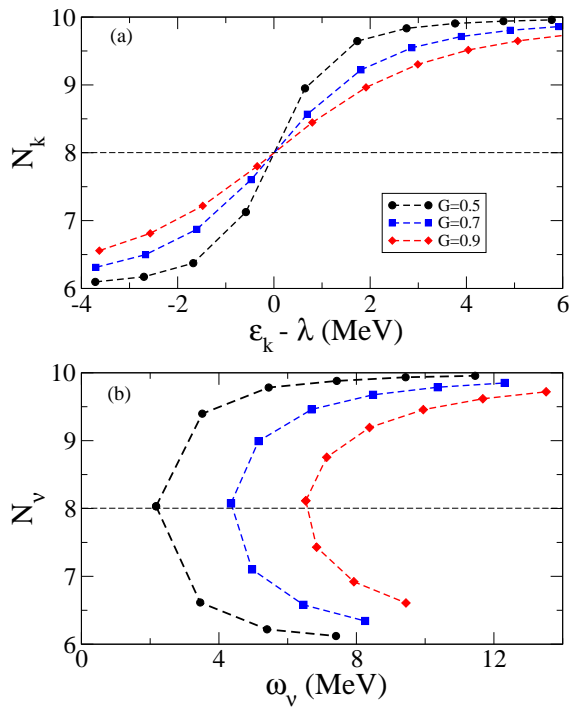}  
\caption{ (color online) Bottom: mean number of particles $N_\nu$ of the QRPA states as a function of their
excitation energy $\hbar \omega_\nu$. Top: mean number of particles $N_k$ of $2$QP states as a function of $(\epsilon_k - \lambda)$, where $\lambda$ is the Fermi energy.}
\label{fig4:pair} 
\end{figure}

At this
stage, it is most likely to conjecture that the failure of QRPA to reproduce
two-particle transfer processes stems from the mixing of systems with different
particle numbers. The QRPA approach implicitly assumes that the states $| \nu
\rangle$ are relatively good approximations for the eigenstates of systems with
$N+2$ (or $N-2$) particles.  As an illustration of the correctness of this assumption, the
mean number of particles $N_\nu = \langle \nu |\hat N | \nu\rangle$ is displayed
in the bottom panel of Fig.  \ref{fig4:pair} as a function of the excitation energy $\omega_\nu$ . 
$ N_\nu$ has been estimated using the quasi-boson approximation leading to:
 \begin{equation}
 N_\nu= \sum_i 2(U_i^2-V_i^2)(X_i^{\nu2}+Y_i^{\nu2}) +\langle \hat N \rangle,
  \end{equation}
where $\langle \hat N \rangle$ is the number of particle in the QP vacuum. 

In the same figure,  the mean-particle number of the two quasi-particle (2QP) excited 
states $| k\rangle$ as a function of $(\epsilon_k - \lambda)$ is also shown, where $\lambda$ is the Fermi energy.
The 2QP states are defined through:
  \begin{eqnarray}
| k \rangle & = & \alpha^\dagger_k \alpha^\dagger_{\bar k} | 0, QP \rangle,  
\end{eqnarray}
where the ground state has $N$ particles in average. The mean particle number 
in $| k \rangle$ is given by:
\begin{eqnarray}
N_k = \langle k |\hat N | k \rangle& = & \langle \hat N \rangle + 2 U^2_k - 2 V^2_k.
\end{eqnarray} 
This expression as well as the illustration in Figure \ref{fig4:pair} clearly
shows that the 2QP states will be close to a state with $N+2$ (resp. $N-2$)
particles only if $U_k \rightarrow 1$, i.e. well above the Fermi energy $\lambda$ (resp.
$U_k \rightarrow 0$, i.e. well below the Fermi energy), but will be a bad
approximation if the 2QP state involves  single-particle state in the vicinity of
$\lambda$. Consequently, QRPA states will also suffer from the same problems if
the state is constructed from 2QP states that are close to the Fermi energy.  While the QRPA results 
are in a reasonable agreement with the exact case, the effect of particle number conservation on the pair transfer
is largely uncontrolled within QRPA.  This is anticipated to be especially crucial in exotic nuclei as the 
level spacing is reduced closed to the drip line. 

\section{Improved treatment of pair transfer in superfluid systems}

The effect of particle number conservation on pair transfer from ground state to ground state
has been already studied in Ref. \cite{Gra12}. It has been empirically found that the breaking of $U(1)$ 
symmetry has a rather small impact on the estimated probabilities. As a further illustration, we show in Figure 
\ref{fig2bis:pair} estimations of transfer probabilities with equation (\ref{eq:add}) 
using the ground state quasi-particle vacua projected either on $N$ or on $N+2$ particle numbers 
(see Ref. \cite{Gra12} for technical details).  As seen in Fig.  \ref{fig2bis:pair}, the BCS approach reduces to HF below 
the threshold and is not able to reproduce the transfer probability at low G. The projection after variation 
(open square) obviously does not cure this problem but considerably improves the probability above the threshold
especially in the strong coupling regime. Note that
As a conclusion, the HFB and/or BCS approach augmented by an a posteriori projection 
is already very good to describe GS to GS transition in the superfluid regime. Therefore, in the 
following we will concentrate the discussion on excited state where it is necessary to go beyond HFB.    

From previous discussion, we are facing the following dilemma: to describe the
physics of pairing and in particular the gap in energy between the ground state
and the first excited state in a superfluid system, it is necessary to break the
$U(1)$ symmetry. On the other hand, this symmetry breaking seems to be at the
origin of some discrepancies between QRPA and exact pair transfer probabilities. 

The most direct extension of the projection technique 
to estimate transition from ground state to ground state would be to directly 
estimate the transition from the BCS/HFB ground state projected onto $N$ particle to the 
QRPA eigenstates projected onto $N+2$ particles. This has however two disadvantages (i) in practice
we found that the pair transfer probabilities that are much smaller than the exact ones (ii) QRPA states are not anymore orthogonal after projection leading to some difficulties to interpret the probabilities themselves.  

Alternatively, one can try to develop a RPA like approach directly in the space of projected 2QP states.  
Following the Tamm-Dancoff approximation spirit,
a set of states $| \Phi_k \rangle $ defined through:
\begin{eqnarray}
| \Phi_k \rangle & = & \hat P_{N+2} \alpha^\dagger_k \alpha^\dagger_{\bar k} | 0 , QP \rangle .
\end{eqnarray} 
are introduced, 
where $\hat P_{N+2} $ is the projector on $N+2$ particles \cite{Bla86}. Then,
excited states of the system with $N+2$ particles decompose as 
\begin{eqnarray}
| \nu , N+2 \rangle = \sum_k X^\nu_k | \Phi_k \rangle. \label{eq:eigenproj}
\end{eqnarray} 
This strategy has been analyzed in Ref. \cite{Kyo90} as well as its RPA generalization following Refs. 
\cite{Sch89} (see also \cite{Rad05}). 

A proper
description would require a full Projected QRPA calculation 
whose practical implementation
would be rather cumbersome, especially in realistic calculations. A simpler approach is
to introduce what could be considered as a projected version
of a two quasiparticle  Tamm-Dancoff approximation. Again, it should be mentioned that contrary 
to standard TDA, states $| \Phi_k \rangle$ are neither normalized nor orthogonal with each others.
Therefore, a special attention has to be paid when formulating the approach. In practice, this implies 
to diagonalize the overlap matrix $O_{kl} = \langle \Phi_k | \Phi_l
\rangle$ prior to write the TDA eigenvalue problem. A practical method, 
has been proposed in ref. \cite{Kyo90}
to obtain the TDA equation in the projected space. Following \cite{Kyo90}, the excited states are written in terms 
of new states  
\begin{eqnarray}
| \nu , N+2 \rangle = \sum_k X^\nu_k | (\Phi_k) \rangle. \label{eq:eigenproj2}
\end{eqnarray}    
States $| (\Phi_k) \rangle$ are defined though:
\begin{eqnarray}
| (\Phi_k) \rangle &=& | \Phi_k \rangle - \langle 0_{\rm N+2} | \Phi_k \rangle, \label{eq:ortho}
\end{eqnarray}
where $| 0_{\rm N+2} \rangle$ correspond here to the approximated ground state with $N+2$ particles. While in the original article, this ground state was anticipated to be obtained with a Variation After Projection (VAP) procedure, below it is simply taken as $| 0_{\rm N+2} \rangle \simeq 
 \hat P_{N+2} | 0, {\rm QP} \rangle$. Then the TDA eigen-equation is solved in the space of  $| (\Phi_k) \rangle$ states. In the following, this approach is referred to Projected two Quasi-Particle TDA (P-QTDA).

\begin{figure}[htbp] 
\includegraphics[width=8cm.]{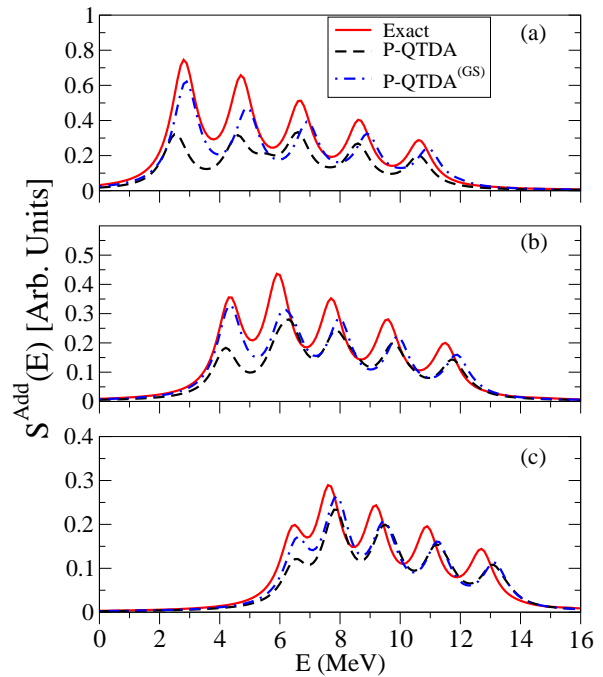} 
\caption{ (color online)  (color online) Same as figure \ref{fig1:pair} for (a) $G=0.5$ MeV, (b) $G=0.7$ MeV and (c) $G=0.9$ MeV.  
The exact result (red solid line) is compared to the   P-QTDA (black dashed line)
and P-QTDA$^{\rm (GS)}$ (blue dot-dashed line). }
\label{fig:ptda} 
\end{figure}  

\begin{figure}[htbp] 
\includegraphics[width=8cm.]{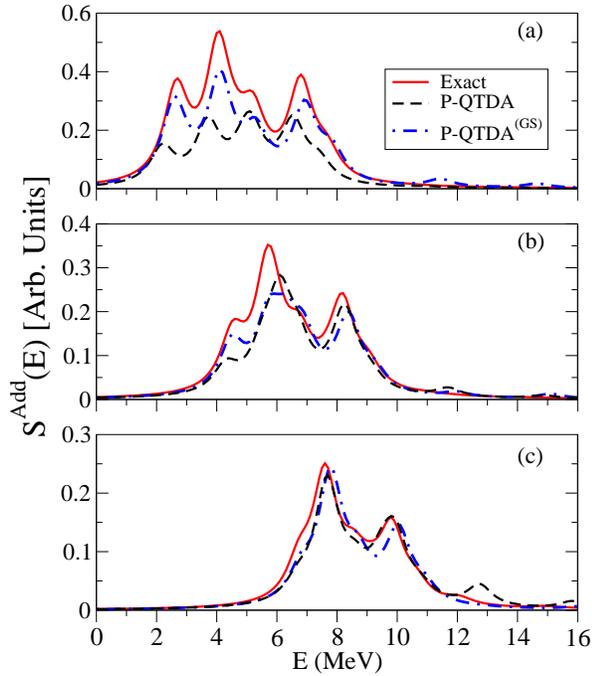} 
\caption{ (color online) Same as figure \ref{fig:ptda} for the case of random level spacing.}
\label{fig:ptdaran} 
\end{figure}  

Consistently with the present approach, addition pair transfer probabilities are computed
using the expectation value of the transition operator between the
quasi-particle vacuum projected on particle $N$ and the excited states
(\ref{eq:eigenproj2}):
 \begin{eqnarray}
P^{\rm Add} = \left | \sum_\nu  X^\nu_k  \langle 0, {\rm QP} |\hat P_N  \hat T  \hat P_{N+2} | (\Phi_k) \rangle \right|^2 \label{eq:piproj} .
\end{eqnarray}
The projection onto different particle numbers in the expression does not induce extra 
difficulty using the fact that 
\begin{eqnarray}
\langle 0, {\rm QP} |\hat P_N  \hat T  \hat P_{N+2} | \Phi_k \rangle &=& \langle 0, {\rm QP} |\hat T  \hat P_{N+2} | \Phi_k \rangle \nonumber \\
&=& \langle 0, {\rm QP} |\hat P_N  \hat T | \Phi_k \rangle .  \nonumber
\end{eqnarray}
Therefore expectation values entering in expression (\ref{eq:piproj}) can be performed using standard projections techniques.
In practice, the projection on particle number is made by discretizing the gauge angle integration using the Fomenko approach with 199 points 
\cite{Fom70,Ben09}. Useful expression to express the
overlaps as well as the Hamiltonian expectation value in a projected basis can
be found in the appendix of Ref. \cite{Lac12}.

Illustration of the method proposed in ref. \cite{Kyo90} applied to pair transfer 
are shown in Fig. \ref{fig:ptda} and Fig. \ref{fig:ptdaran} respectively for the equidistant and random single-particle level spacing. 
These figures clearly demonstrate that the projected TDA approach provides a much better description of the energy peak positions 
of the excited states in the system with $N+2$ particles. However, probabilities of transfer are underestimated especially as $G$ 
decreases.   
    
Further improvements can a priori be made by including more correlations in the ground state $| 0_{\rm N+2} \rangle$. Here, we simply 
used a Projection After Variation (PAV) approach to approximate this state. A better treatment would be to perform a VAP as originally proposed
in ref. \cite{Kyo90}. Further correlations, like correlations induced by the coupling to 4QP states might be included using Projected QRPA 
instead of Projected TDA approach. The Projected QRPA approach has also been formulated previously (section III-B of Ref. \cite{Kyo90}).  
In that case, not only the projected 2QP states but also the projected 4QP states should be explicitly introduced.  

The use of VAP and/or the introduction of 4QP, although possible in the present model \cite{San08,Hup11,Lac12}, will considerably 
increase the complexity of the approach in realistic situations (see for instance \cite{ Rod07,Rod10, Hup12} for the VAP). Below, we
propose a simpler method inspired from P-QTDA and able to grasp part of the ground state correlations
without increasing the numerical complexity. 
  
Contrary to standard linear response theory based on HF (resp. HFB), the ground state projected onto good particle number is not 
orthogonal to the particle-hole (resp. 2QP) excited states. Similarly, the projected 2QP states are themselves not orthogonal 
to the projected 4QP or higher order excitations. At first glance, this might be seen as a disadvantage compared to RPA or QRPA since additional 
orthonormalization is required. On the contrary, one might take advantage of the presence of higher order components to improve the description of the GS itself. 

The aim of the P-QTDA approach was to describe excitations with respect to the projected ground state. 
Then, the latter state has been naturally removed before solving the eigenvalues equation (Eq. (\ref{eq:ortho})).
Let us assume that we restart from expression (\ref{eq:eigenproj}) where the projected mean-field is also 
included in the sum ( with the convention that 
$| \Phi_0 \rangle  =  \hat P_{N+2} | 0 , QP \rangle$). Then, 
coefficients $X^\nu_k$ can be obtained by diagonalizing the hamiltonian in the 
$\{ \Phi_k \}$ (with proper treatment of the non-orthonormality of the states). This direct approach, called below 
P-QTDA$^{\rm (GS)}$, 
not only provides a way to get excited states but might also improve the description of the ground state 
itself. This is indeed what we observed empirically. For instance, for $G=0.5$ MeV and equidistant level spacing, 
the new ground state has an energy $300$ keV lower than the energy of the original
projected mean-field ground state. The difference reduces has $G$ increase.  At $G=0.9$ MeV the reduction 
is only $30$ keV. 
For completeness, the probability to transfer from GS to GS within the P-QTDA$^{\rm (GS)}$ is also shown in Figure \ref{fig2bis:pair}.
This probability is slightly improved compared to the BCS case below the BCS threshold while it follows the PAV results above.  

It turns out, that this approach improves the description of pair transfer from GS to excited states. 
Results of the P-QTDA$^{\rm (GS)}$ method are presented in Fig. (\ref{fig:ptda}) and (\ref{fig:ptdaran}) by dashed line.
A clear improvement is observed especially at low excitations and low $G$. 
The remaining difference with the exact solution is acceptable in view of the 
simplified approach presented here.

\begin{figure}[htbp] 
\includegraphics[width=8cm.]{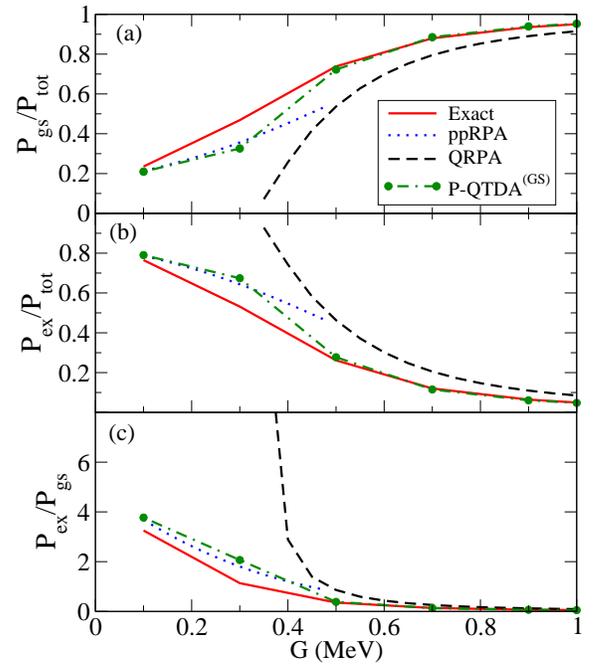} 
\caption{ (color online) Ratios of probabilities estimated with different theories for the equidistant level case. 
The exact (red solid line), QRPA (black dashed line) and P-QTDA (green filled circles) are shown. The ppRPA is also shown
(dotted line) up to $G=0.48$ MeV.}
\label{fig7:pair} 
\end{figure}  
To further quantify the predictive power of the P-QTDA$^{\rm (GS)}$ method, some ratios of pair transfer probabilities 
are shown in Figure \ref{fig7:pair}. In this figure, $P_{\rm gs}$, $P_{\rm ex}$ and $P_{\rm tot}$ 
correspond respectively to the probability to transfer to the ground state, the sum of probabilities to transfer 
to any excited states, while $P_{\rm tot} = P_{\rm gs} + P_{\rm ex}$. Below the pairing threshold, the 
P-QTDA$^{\rm (GS)}$ reduces to a ppTDA$^{\rm (GS)}$, where the diagonalization is made in a reduced space of Slater 
determinant. In that case, the result are of the same quality as for the ppRPA.
This figure clearly confirms that while QRPA is rather far from 
the exact results in the superfluid phase, especially in the vicinity of the BCS threshold, 
the projected theory provides a much better reproduction of 
 ground state to excited state pair transfer probabilities.  
Moreover,theoretical predictions  based on the QRPA \cite{For02} might significantly overestimate  the GPV cross section,
 as suggested also in \cite{Mou11}, especially in the vicinity of the normal-superfluid transition.

\section{Conclusions}

In this work, the QRPA description of the two-particle transfer mechanism is
tested against the exact solution in the Richardson model 
for several conditions of pairing interaction strength and level spacing. 
It is seen that both ppRPA in the normal phase and QRPA in the superfluid 
region are able to grasp the gross feature of the pair transfer process. 
However some differences are observed. At variance with other kind of resonance states mainly build in terms 
of particle-hole excitations (see for instance Fig. 1 of Ref. \cite{Ter08})
the particle number conservation seems to play an important role when the particle number change 
during the physical process under interest. A method is proposed here to improve 
the description of pair transfer in finite superfluid systems. 
The new method is based on the 
direct diagonalization of the Hamiltonian in a reduced space of the 
projected ground state plus two quasi-particle states. This theory improves 
considerably the description of the pair transfer process. 
On the practical side, the P-QTDA$^{\rm (GS)}$ requires only to solve the BCS or HFB
problem in the initial nucleus and, except the additional numerical cost of projection, it 
does not need more effort than the original QRPA.
Work is in progress to apply it to nuclear transfer reactions.
     
\begin{acknowledgments} 
D.L. gratefully acknowledges IPN Orsay for the support and warm hospitality extended to him. 
We would like to thank M. Grasso for discussions at different stage of this work.
\end{acknowledgments}

\end{document}